\documentclass[preprint,prb,superscriptaddress,showpacs]{revtex4}

\usepackage{amssymb, amsmath}

\usepackage{graphicx}
\usepackage{color}

\begin{document}

\title{Complex itinerant ferromagnetism in noncentrosymmetric Cr$_{11}$Ge$_{19}$}

\author{N. J. Ghimire}

%\email{nghimire@utk.edu}

\affiliation{Department of Physics and Astronomy, The University of Tennessee, Knoxville, Tennessee 37996, USA}

\affiliation{Materials Science and Technology Division, Oak Ridge National Laboratory, Oak Ridge, Tennessee 37831, USA}

\author{M. A. McGuire}

%\email{mmcguire@ornl.gov}

\affiliation{Materials Science and Technology Division, Oak Ridge National Laboratory, Oak Ridge, Tennessee 37831, USA}

\author{D. S. Parker}

%\email{parkerds@ornl.gov}

\affiliation{Materials Science and Technology Division, Oak Ridge National Laboratory, Oak Ridge, Tennessee 37831, USA}

\author{B. C. Sales}

%\email{salesbc@ornl.gov}

\affiliation{Materials Science and Technology Division, Oak Ridge National Laboratory, Oak Ridge, Tennessee 37831, USA}

\author{J.-Q. Yan}

%\email{jyan6@utk.edu}

\affiliation{Materials Science and Technology Division, Oak Ridge National Laboratory, Oak Ridge, Tennessee 37831, USA}
\affiliation{Department of Materials Science and Engineering, The University of Tennessee, Knoxville, Tennessee, 37996, USA}

\author{V. Keppens}

%\email{vkeppens@utk.edu}

\affiliation{Department of Materials Science and Engineering, The University of Tennessee, Knoxville, Tennessee, 37996, USA}

\author{M. Koehler}

%\email{mrkoehler@gmail.com}

\affiliation{Department of Materials Science and Engineering, The University of Tennessee, Knoxville, Tennessee, 37996, USA}

\author{R. M. Latture}

%\email{rlatture@utk.edu}

\affiliation{Department of Materials Science and Engineering, The University of Tennessee, Knoxville, Tennessee, 37996, USA}

\author{D. Mandrus}

%\email{dmandrus@utk.edu}

\affiliation{Department of Physics and Astronomy, The University of Tennessee, Knoxville, Tennessee 37996, USA}
\affiliation{Materials Science and Technology Division, Oak Ridge National Laboratory, Oak Ridge, Tennessee 37831, USA}
\affiliation{Department of Materials Science and Engineering, The University of Tennessee, Knoxville, Tennessee, 37996, USA}

\date{\today}

\begin{abstract}

The noncentrosymmetric ferromagnet Cr$_{11}$Ge$_{19}$ has been investigated by electrical transport, AC and DC magnetization, heat capacity, x-ray diffraction, resonant ultrasound spectroscopy, and first principles electronic structure calculations. Complex itinerant ferromagnetism in this material is indicated by nonlinearity in conventional Arrott plots, unusual behavior of AC susceptibility, and a weak heat capacity anomaly near the Curie temperature (88 K). The inclusion of spin wave excitations was found to be important in modeling the low temperature heat capacity. The temperature dependence of the elastic moduli and lattice constants, including negative thermal expansion along the c axis at low temperatures, indicates strong magneto-elastic coupling in this system. Calculations show strong evidence for itinerant ferromagnetism and suggest a noncollinear ground state may be expected.

\end{abstract}
\pacs{75.30.Kz, 75.10.-b}

\maketitle

\section{Introduction}

Itinerant ferromagnets crystallizing in noncentrosymmetric space groups have attracted much attention recently. The key factor is that the lack of an inversion center in the crystal lattice means that Dzyaloshinsky-Moriya (DM) spin-orbit interactions are allowed. \cite{Dzyaloshinsky1958,Moriya1960} These interactions add a term to the free energy $\overrightarrow{m}\cdot\nabla \times \overrightarrow{m}$ favoring perpendicular orientation of the spins. In metallic systems the DM term can lead to helimagnetism \cite{Bak1980a,Nakanishi1980b} and complex spin textures resembling liquid crystal phases when the helimagnetism is destabilized. \cite{Fischer2008a,Muhlbauer2009,Ho2010b} MnSi is the most heavily studied system in this class of materials. MnSi orders at 29.5 K forming a long period helimagnet with a wavelength $\lambda_{h}\sim$ 180 \AA\ weakly pinned along the $<$111$>$ direction. It crystallizes in the noncentrosymmetric space group $P2_{1}3$ with the B20 structure. At low temperatures and in applied magnetic fields above 6 kOe a field polarized phase appears. Just below the transition temperature and for magnetic fields applied along $<$100$>$ a phase known as the \textit{A} phase is stabilized. Recently this phase has been identified as a skyrmion lattice. \cite{Pfleiderer2009a, Muhlbauer2009} Skyrmions are a direct result of DM interactions allowed in the crystal lattice with no inversion symmetry. Similar spin textures have been observed in two other materials so far, FeGe \cite{Uchida2006b} and Fe$_{1-x}$Co$_{x}$Si, \cite{Munzer2010} both having the same structure as MnSi. Cr$_{1/3}$NbS$_{2}$ is another example where the DM interaction is responsible for the helimagnetism. \cite{Miyadai1983,Moriya1982} In this material the period of the helix is $\sim 480$ \AA, and the DM interaction is thought to be stabilized by the lack of an inversion center between the two chromium atoms along the c axis as it crystallizes in a noncentrosymmetric space group $P$6$_{3}$22. Recently, Lorentz microscopy and small-angle electron diffraction studies showed an emergence in the applied magnetic field of a periodic and nonlinear magnetic order called a chiral magnetic soliton lattice in addition to the the zero-field chiral helimagnetic structure. \cite{Togawa2012}

Indeed, there are many other ferromagnets that crystallize in space groups lacking an inversion center. However, until now, both the theoretical and experimental studies have been concentrated mostly within the materials with the cubic B20 crystal structure.  Thus investigation of noncentrosymmetric magnets with different structures is important for understanding the consequences of DM interactions in a wide variety of compounds. Cr$_{11}$Ge$_{19}$ is such a material\cite{Vollenkle1967,Zagryazhskii1968,Kolenda1980}. It crystallizes in the noncentrosymmetric space group $P\overline{4}n$2 belonging to the point group $D^{8}_{2d}$ and orders ferromagnetically below about 90 K. Interestingly, $D_{2d}$ is one of the crystallographic classes in which a ferromagnet is expected to have a thermodynamically stable magnetic vortex phase in a certain range of applied magnetic field.  \cite{Bogdanov1989a, Bogdanov1989b} This phase is reminiscent of an Abrikosov vortex lattice in a type II superconductor. The possibility of such a structure was also predicted \cite{Bogdanov1994} for MnSi, FeGe, Fe$_{x}$Co$_{1-x}$Si, and Co$_{x}$Mn$_{1-x}$Si. Skyrmion lattices have since been identified in the first three of these compounds, as stated above. Relatively few studies have appeared on Cr$_{11}$Ge$_{19}$. In early work, Zagryazhskii \emph{et al.} \cite{Zagryazhskii1968} reported Cr$_{11}$Ge$_{19}$ to be a semimetallic ferromagnet with a transition temperature of  $\sim$ 86 K. Intriguingly, they point out the lack of a lambda anomaly at the ferromagnetic transition temperature in their specific heat measurements. A linear muffin tin orbital (LMTO) calculation of electronic density of states \cite{Pecheur1997} indicated it to be a low moment itinerant ferromagnet. A study of thermoelectric properties on a single crystal above room temperature reported the material to have a metallic behavior with dominant p-type conductivity and a relatively low resistivity. \cite{Caillat1997}

In this paper we report magnetization, transport, and thermodynamic properties of Cr$_{11}$Ge$_{19}$ together with results obtained from electronic structure calculations.  Both the experimental results and calculations indicate that Cr$_{11}$Ge$_{19}$ is a good example of an itinerant electron ferromagnet, with signatures of both spin wave excitations and magnetic fluctuations apparent in the data. Although no direct evidence for a helimagnetic or other exotic magnetic ground state has been found in the polycrystalline samples we have studied, the behavior of this material is unusual in several respects and deserves further study in single crystal form.

\section{Experimental details}

Polycrystalline samples were prepared and studied. Stoichiometric amounts of high purity Cr pieces (99.999 $\%$) and Ge pieces (99.9999 $\%$) were arc-melted in an argon atmosphere. The resulting ingot was then sealed in a quartz tube and annealed at 900 $^{o}$C for one week. The annealed ingot was then ground into fine powder inside a He-filled glove box and pressed into a pellet, which was again sealed in an evacuated quartz tube and annealed at 900 $^{o}$C for another week.

Single crystal growth was also attempted using two different techniques: a flux method using Ge as a self flux and a modified Bridgman method. Both growths were carried out using a molar ratio of Cr : Ge = 20 : 80 of the starting materials. In the flux method, a total charge consisting of 7 g of  Ge pieces (99.9999 $\%$  pure) and Cr powders (99.99 $\%$ pure)  were loaded in a 5 ml alumina crucible.  A catch crucible containing quartz wool was mounted on top of a growth crucible and both were sealed in a silica ampoule under  vacuum.  The sealed ampoule was  heated to 1100 $^{o}$C over 10 hours and homogenized at 1100 $^{o}$C for 30 hours, furnace cooled to 1000 $^{o}$C  and then slowly cooled to 910 $^{o}$C  at the rate of 2 $^{o}$C per hour. Once the furnace reached 910 $^{o}$C,  the excess flux was decanted from the crystals.  Single crystals with cubic shape and of an average dimension of about 0.5 mm were obtained. The so called modified Bridgman method was employed by first melting a total charge of 10 g of Ge pieces (99.9999 $\%$ pure) and Cr pieces (99.999 $\%$) in an argon atmosphere. The arc-melted ingot was broken into pieces and loaded into a well-cleaned quartz tube of 14 mm inner diameter with  a pointed bottom forming a Bridgman crucible. The tube was placed in an upright position inside a box furnace and first heated to 1100 $^{o}$C over 10 hours and homogenized for 30 hours. It was then cooled quickly to 1000 $^{o}$C  and then slowly  cooled to 900 $^{o}$C  at the rate of 2 $^{o}$C  per hour which was subsequently furnace-cooled to room temperature. Tiny cube-shaped crystals with a typical dimension of  0.1 mm were always observed in the middle of the resulting boule.

Room temperature x-ray diffraction on powders  from pulverized single crystals confirmed single phase for the crystals obtained in both techniques. The atomic ratio was studied using a Hitachi bench-top scanning electron microscope (SEM) with a Bruker energy dispersive x-ray spectrometer (EDS). The atomic percentages of Cr and Ge observed are 42 at.$\%$ and 58 at.$\%$, respectively, which is within the expected uncertainty for standardless measurements on irregular surfaces. Crystals obtained from modified Bridgman method  were too small for convenient characterization,  while the relatively larger crystals grown by flux method typically had some residual Ge flux on the crystal surface. Therefore in this study we characterized polycrystalline material, which was $\geq$98\% pure based on powder diffraction and EDS measurements.

 All the measurements were carried out in pieces cut from the same compact polycrystalline pellet that was determined to be 82$\%$ of the theoretical density.  X-ray powder diffraction was performed at room temperature using a PANalytical X'Pert powder diffractometer for phase identification and structural refinement.  X-ray powder patterns were also obtained every 20 K on cooling from room temperature down to 11 K. DC magnetization measurements were performed using a Quantum Design magnetic property measurement system (MPMS). AC susceptibility, specific heat, and resistivity measurements were conducted in a Quantum Design physical property measurement system (PPMS). AC susceptibility was measured by using a drive coil frequency of 85 Hz and an excitation field of 10 Oe at different applied DC magnetic fields from 0 to 10 kOe. Specific heat measurements were performed on a small piece of 27.5 mg. Resistivity was measured using platinum wires and Epotek H20E silver epoxy in a four-wire configuration on a 1.9$\times$1.2$\times$ 1.4 mm$^3$ rectangular bar.  The temperature dependence of the elastic moduli was obtained using resonant ultrasound spectroscopy (RUS) on a 1.261 x 2.122 x 3.899 mm$^{3}$ polycrystalline pellet using a custom designed probe inserted into a Quantum Design Versalab.\cite{Migliori2005}

\section{Results and discussion}

\subsection{Crystal chemistry}
\begin{figure}[h]
\begin{center}
\includegraphics[scale=2]{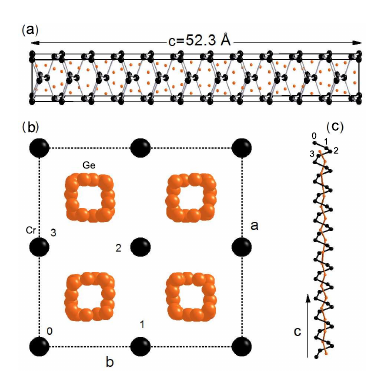}
\caption{\label{1}(Color online) The tetragonal Cr$_{11}$Ge$_{19}$ structure. (a) Arrangement of Cr and Ge atoms in the complex Nowotny chimney ladder structure emphasizing the long c axis (52.321 \AA). (b) A view down the c axis. One turn of the Cr helix is emphasized on moving from 0, 1, 2, and 3 counterclockwise. (c) A perpendicular view showing a Ge helix within a Cr helix. The Cr atoms are shown as black (larger) balls, and the Ge atoms are shown as orange (smaller) balls.}
\end{center}
\end{figure}
\begin{figure}[h]
\begin{center}
\includegraphics[scale=2.0]{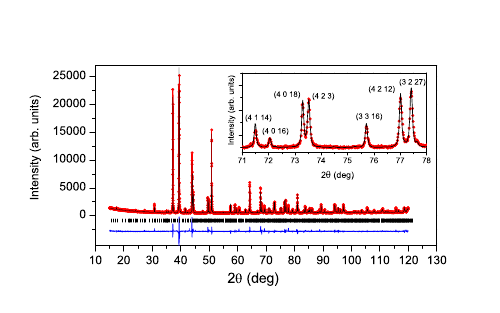}
\caption{\label{2}(Color online) Rietveld refinement of x-ray powder pattern of Cr$_{11}$Ge$_{19}$ collected at room temperature.}
\end{center}
\end{figure}

Cr$_{11}$Ge$_{19}$ crystallizes in the Mn$_{11}$Si$_{19}$ structure type in a family of compounds known as Nowotny chimney ladders (NCLs). These are a series of intermetallic compounds with composition $T_{n}X_{m}$, where 2$>$m/n$>$1.25.\cite{Caillat1997} Here \emph{T} is a transition metal element and \emph{X} is a main group metal. These compounds have a complex structure in which \emph{T} atoms form 4-fold helices inside of which \emph{X} atoms form separate helices.\cite{Fredrickson2004a} NCLs have been found to follow the 14 electron rule, according to which a NCL compound having 14 valence electrons (VEL) per main group metal atom should be semiconductor and one with VEL less than 14 should be metal.\cite{Caillat1997} The rule holds for Cr$_{11}$Ge$_{19}$ as it has total of 12.9 valence electrons per Ge atom and is known to have metallic behavior. Figure \ref{1} shows the structure of Cr$_{11}$Ge$_{19}$. It has a very long c axis (52.321 \AA) as shown in Fig. \ref{1}(a). Figure \ref{1}(b) shows the view down the c axis. The Cr atoms (black) form helices, shaped like chimneys, within which the helices of Ge atoms (orange) reside. In Fig. \ref{1}(c) we emphasize the helices of Cr and Ge. The Cr-Cr distance is the shortest along the helix (3.124 - 3.138 \AA), \cite{Vollenkle1967} and hence, substantial Cr-Cr interaction can be expected in the direction of the helix. However, the structure is much more complex due to the presence of large number of atoms (120) in the unit cell. It has 12 inequivalent Cr sites and 10 inequivalent Ge sites. We used the reported structure\cite{Vollenkle1967} for the Rietveld refinement of the room temperature x-ray powder pattern. The fit is reasonably good considering the complex structure as shown in Fig. \ref{2}. Atomic positions and occupancies were not refined because of the difficulty introduced by the large number of atoms in the unit cell. The inset in Fig. \ref{2} shows a magnified part of the fit at higher angles in which indexed peaks are seen more clearly. Lattice constants obtained from the fit are a = 5.805 \AA\ and c = 52.321 \AA, which are in good agreement with previously reported values. \cite{Zagryazhskii1968,Pecheur1997}

\subsection{DC magnetization}

\begin{figure}[h]
\begin{center}
\includegraphics[scale=2]{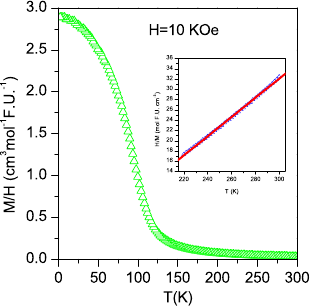}
\caption{\label{3}(Color online) $M/H$ as a function of temperature measured at an applied field of $H$ = 10 kOe. Inset shows the fit to the Curie-Weiss law.}
\end{center}
\end{figure}
Figure \ref{3} shows the temperature dependence of magnetization of Cr$_{11}$Ge$_{19}$ in an applied field of 10 kOe. As the sample is cooled the transition from a paramagnetic state to an ordered ferromagnetic state is clearly visible. The inset shows the Curie-Weiss fit of $\chi^{-1}$ = $(T-\theta_{CW})/C$ to the high temperature part of the inverse susceptibility above 220 K. The parameters obtained are Curie constant $C$ = 5.34 \emph{K cm$^{3}$ mol$^{-1}$ F.U.$^{-1}$} and the Curie-Weiss temperature $\theta_{CW}$ = 128.6 K. The effective moment per mole of chromium atom $p_{eff}$ calculated from the Curie constant is 1.97 $\mu_{B}$.

\begin{figure}[h]
\begin{center}
\includegraphics[scale=1.5]{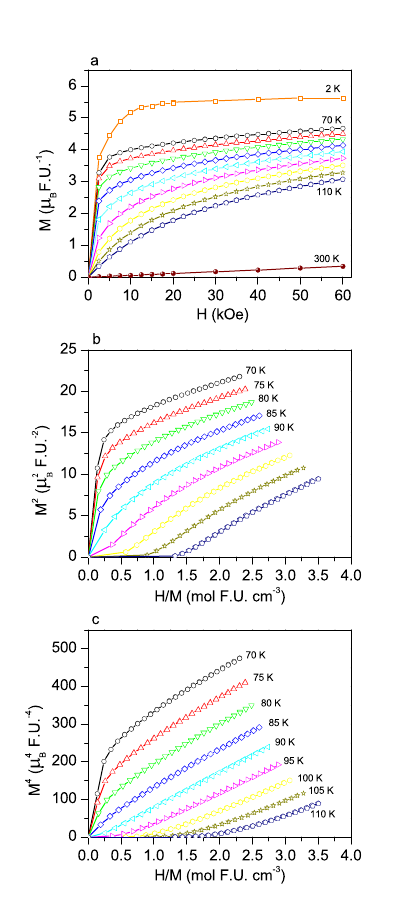}
\caption{\label{4}(Color online) (a) $M$ versus $H$ for $Cr_{11}Ge_{19}$ at indicated temperatures. The plots from 70 K to 110 K are in the interval of every 5 K. (b) $M^{2}$ versus $H/M$ (Arrott plot) and (c) $M^{4}$ versus $H/M$ for $Cr_{11}Ge_{19}$ at indicated temperatures.}
\end{center}
\end{figure}

Figure \ref{4}(a) shows the magnetization $M$ of Cr$_{11}$Ge$_{19}$ as a function of field at several temperatures. At low temperatures $M$ saturates above 20 kOe. The saturation is suppressed with increasing temperature, and $M$ versus $H$ becomes a straight line at higher temperatures. The saturated magnetic moment obtained in the ordered state is 0.49 $\mu_{B}$/$Cr$. Within the Stoner model, itinerant ferromagnets obey the relation
\begin{equation}\label{eq1}
    M(H,T)^{2}=-\frac{A}{B}+\frac{1}{B}(\frac{H}{M(H,T)}),
\end{equation}
where $A$ and $B$ are independent of $H$. \cite{Wohlfarth1977,Wohlfarth1968,Edwards1968} $A$ is a temperature dependent term and vanishes at $T_{C}$. This should give straight lines on an Arrott plot, \cite{Arrott1967} $M^{2}(H,T)$ versus $H/M(H,T)$, with a straight line passing through the origin at the transition temperature . In Fig. \ref{4}(b) we show Arrott plots for Cr$_{11}$Ge$_{19}$. These Arrott plots are not perfectly straight lines as expected and observed in itinerant ferromagnets like ZrZn$_{2}$, Ni$_{3}$Al, and NiPt alloys. \cite{Wohlfarth1977} However, the Arrott plots for Cr$_{11}$Ge$_{19}$ are similar to those observed in the case of MnSi \cite{Chattopadhyay2009} and the layered itinerant ferromagnet LaCoAsO.\cite{Ohta2009} Such behavior was explained by Takahashi,\cite{Takahashi1986} who in his theory added zero point local spin fluctuations, which were previously neglected. This theory predicts
\begin{equation}\label{eq2}
    h=[\frac{T_{A}}{3}(2+\sqrt{5})T_{c}]^{2}m^{5},
\end{equation}
where h = $2\mu_{B}$H and m = $2M(T)/N_{o}$ magnetization per magnetic site. The parameter $T_{A}$ characterizes the dispersion of the static magnetic susceptibility in wave-vector (\emph{q}) space. From Eq. (2) it can be seen that $M^{4}$ versus $H/M$ should be a straight line at $T_{C}$ . Such a linear relation has been confirmed in MnSi and Fe$_{1-x}$Co$_{x}$Si. \cite{Takahashi1986,Shimizu1990} Figure \ref{4}(c) shows the $M^{4}$ versus $H/M$ curve of Cr$_{11}$Ge$_{19}$ which shows straight line behavior. The plots are almost a straight line between 85 and 90 K. We estimate the Curie temperature to be 88 K.
\subsection{AC susceptibility}

\begin{figure}[h]
\begin{center}
\includegraphics[scale=2]{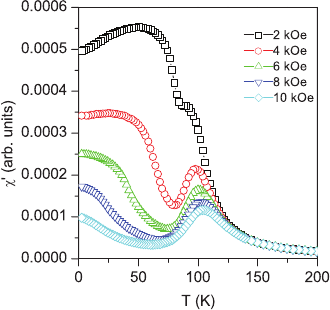}
\caption{\label{5}(Color online) AC susceptibility of Cr$_{11}$Ge$_{19}$ as a function of temperature at indicated applied fields.}
\end{center}
\end{figure}

Figure \ref{5} shows the temperature dependance of the real part of the AC susceptibility measured in the applied DC fields as indicated. At zero applied DC bias field (not shown) the AC susceptibility increases quickly with decreasing temperature in the vicinity of the transition temperature and decreases slightly upon further cooling. The effect of external fields is quite striking. First, the field remarkably suppresses the ac susceptibility. Second, a shoulder appears at lower fields near $T_{C}$ which is defined by two peaks, one sharp peak in the vicinity of the transition temperature and the other broader peak below $T_{C}$. With the increase in field, the peak near $T_{C}$ gets suppressed and shifts slightly towards higher temperature, whereas the broader peak below $T_{C}$ becomes broader and shifts towards lower temperature and is almost completely suppressed at H = 10 kOe. Similar behavior has been observed in a PdMn alloy \cite{SCHo1981}, GdFe$_{2}$Zn$_{20}$ \cite{Vannette2008b}, MnSi\cite{Thessieu1997a} and FeGe.\cite{Wilhelm2011} This AC susceptibility behavior in GdFe$_{2}$Zn$_{20}$ has been interpreted as a manifestation of both the itinerant and the local moments in the material as it contains both 4f (local) and 3d (itinerant) moments. In this material, the peak observed near $T_{C}$ shows the behavior of local moments as observed in CeAgSb$_{2}$, \cite{Vannette2008b} and the broader peak at lower temperature is reminiscent of itinerant behavior as observed in ZnZr$_{2}$.\cite{Vannette2008b} MnSi and FeGe show similar AC susceptibility behavior, but are known to have no local moments. In these later two materials, the \emph{A} phase has been tracked out by AC susceptibility measurements conducted on single crystals. \cite{Thessieu1997a, Wilhelm2011}

\subsection{Heat capacity}

\begin{figure}[h]
\begin{center}
\includegraphics[scale=2]{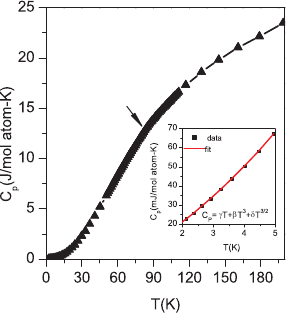}
\caption{\label{6}(Color online) Temperature dependence of molar heat capacity of Cr$_{11}$Ge$_{19}$. Inset shows the low temperature fit of the heat capacity.}
\end{center}
\end{figure}
\begin{figure}[h]
\begin{center}
\includegraphics[scale=2]{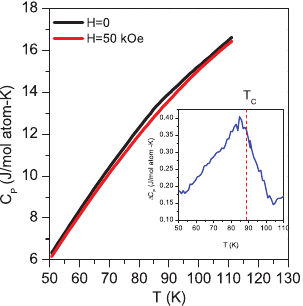}
\caption{\label{7}(Color online) Heat capacity of Cr$_{11}$Ge$_{19}$ in ambient field and 50 kOe. The inset shows $\Delta C_{p}$ = $C_{p}(H=0)$ - \emph{C$_{p}$(H = 50 kOe)}.}
\end{center}
\end{figure}

The molar heat capacity of Cr$_{11}$Ge$_{19}$ from 2 to 200 K is shown in Fig. \ref{6}. There is no obvious lambda anomaly near $T_{C}$, but upon closer inspection $T_{C}$ is marked by a small kink as shown by the arrow in Fig. \ref{6}. Zagryazhskii \emph{et al.} reported that they observed a monotonic increase in the specific heat capacity from 55 to 300 K with no lambda anomaly.\cite{Zagryazhskii1968}  The small kink observed near $T_{C}$ is suppressed upon application of the magnetic field. Figure. \ref{7} shows the specific heat capacity as a function of temperature between 50 and 110 K measured in zero field and at 50 kOe. The inset shows the difference curve obtained by plotting  $\Delta$C$_{p}$ (the difference between  C$_{p}$ measured in a 50 kOe field and in a zero applied field) as a function of temperature giving a clear peak near $T_{C}$. Mohn and Hilscher \cite{Mohn1989} have discussed the influence of spin fluctuations on the specific heat of itinerant ferromagnets. In Stoner theory, the magnetic contribution vanishes above $T_{C}$. In contrast, in systems with spin fluctuations, it is only the macroscopic moment that disappears at $T_{C}$  as spin fluctuations persist above the ordering temperature. The magnetic contribution to the discontinuity in the specific heat at the transition temperature in case of pure single particle excitations is given by $\Delta C_{m}$ = $\frac{M_{o}^{2}}{\chi_{o}T_{C}}$. When spin fluctuations are taken into account, the discontinuity is given by $\Delta$C$_{m}$ = $\frac{M_{o}^{2}}{2\chi_{o}T_{C}}(\frac{1}{2}t_{c}^{4}+t_{c}^{2}+\frac{1}{2})$, where $M_{o}$ is the spontaneous magnetization, $\chi_{o}$ is the initial ferromagnetic susceptibility, $T_{C}$  is the transition temperature, and $t_{c}$ = $T_{C}/T_{C}^{s}$ with $T_{C}^{s}$ being the Curie temperature derived from the pure Stoner type behavior.\cite{Mohn1989} In Cr$_{11}$Ge$_{19}$ the discontinuity in the specific heat at $T_{C}$ calculated for pure Stoner type excitations is 1.2 \emph{J (mol-atom)$^{-1}$ K$^{-1}$}. This value is small enough to explain the absence of a well-defined lambda anomaly in the specific heat capacity near the ferromagnetic transition. But, there is considerable uncertainty in the calculation. The spontaneous magnetization $M_{o}$ is calculated by using the theory applicable for an itinerant ferromagnet \cite{Huber1975} by fitting straight lines obtained at higher fields in the Arrott plot even though the Arrott plot in this material does not behave perfectly as in the case of systems like ZrZn$_{2}$.\cite{Ogawa1967} There can also be appreciable uncertainty introduced by $\chi_{o}$, which might include other components than only the spin susceptibility (e.g., a diamagnetic component). It should be noted that we have not included the contribution due to spin fluctuations in the calculation because of the difficulty introduced by the large number of electrons (1680 per unit cell) in estimating $T_{C}^{s}$ from band structure calculations. Inclusion of spin fluctuations further decreases $\Delta C_{m}$. In case of maximum spin fluctuations $\Delta C_{m}$ is reduced by a factor of 4. Thus, the presence of spin fluctuations in the material reduces the possibility of getting a sizable discontinuity in specific heat at the transition temperature even if some uncertainty might have been introduced in the calculation of $\Delta C_{m}$.

The low temperature specific heat data could not be well modeled by $C_{p} =\gamma T +\beta T^{3}$. This indicates additional excitations may be contributing to the heat capacity at low temperature. Since this material is magnetically ordered below 88 K, magnetic excitations were considered by inclusion of a term in C$_p$ proportional to $T^{\frac{3}{2}}$. \cite{Mohn2006} This resulted in a good fit to the data as shown in the inset of Fig. \ref{6}. The fit yields the electronic heat capacity coefficient $\gamma$ = 7.26 $mJ/molK^{2}$, the phonon specific heat coefficient $\beta$ = 0.06 $mJ/molK^{4}$,  and the magnetic specific heat coefficient $\delta$ = 2.18 $mJ/molK^{5/2}$. The Debye temperature determined from $\beta$ is 319 K. Forcing a fit without the magnetic term gives a much lower value for the fitted Debye temperature ($\sim$ 240 K). Elastic constant data presented below give a Debye temperature of 340 K, further justifying the inclusion of spin excitations in modeling the heat capacity.

\subsection{Resistivity and magnetoresistance}
\begin{figure}[h]
\begin{center}
\includegraphics[scale=2]{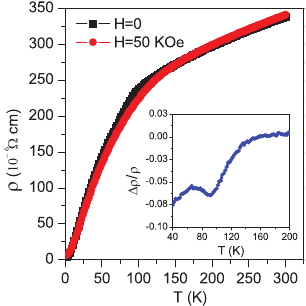}
\caption{\label{8}(Color online) Resistivity of Cr$_{11}$Ge$_{19}$ as a function of temperature. Inset shows the magnetoresistance.}
\end{center}
\end{figure}

Electrical resistivity of Cr$_{11}$Ge$_{19}$ vs temperature is plotted in Fig. \ref{8}. The temperature dependence of the resistivity is metallic over the whole temperature range. A slope change is observed at $\thicksim$ 90 K, which is consistent with a significant loss of spin-disorder scattering upon magnetic ordering. The $T_{C}$ inferred from resistivity is in good agreement with the value of $T_{C}$ obtained from magnetization measurements. The room temperature value of electrical resistivity, 0.35 \emph{m$\Omega$ cm}, is in good agreement with the value (0.345 \emph{m$\Omega$ cm}) reported by Zagryazhskii \emph{et al.} \cite{Zagryazhskii1968} measured on a polycrystal sample and is about a factor of 2 higher than the value reported by Caillat \emph{et al.} \cite{Caillat1997} on a single crystal sample.  The observed excess value of resistivity can be attributed to grain boundary scattering in the polycrystalline sample. The residual resistance ratio ($\rho_{300 K}/\rho_{2 K}$  ) is large, having a value of 89. The inset in Fig. \ref{8} shows the magnetoresistance defined as $\Delta\rho/\rho$,  where $\Delta\rho$ = $\rho_{H}$ - $\rho$ with $\rho_{H}$ and $\rho$  being the resistivity measured at 50 kOe and zero applied magnetic field, respectively. Negative magnetoresistance is observed below 150 K with the largest effect in the vicinity of $T_{C}$ where fluctuations are the strongest.

\subsection{Thermal expansion and elastic moduli}

\begin{figure}[h]
\begin{center}
\includegraphics[scale=2]{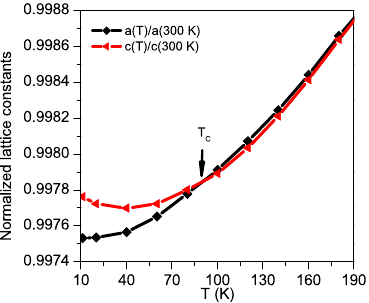}
\caption{\label{9}(Color online) Temperature dependence of lattice constants of Cr$_{11}$Ge$_{19}$. The lattice constants are normalized by dividing with values at 300 K.}
\end{center}
\end{figure}

The temperature dependence of the lattice parameters is plotted in Fig. \ref{9}. Both \emph{a(T)} and \emph{c(T)} are normalized by dividing with the corresponding room temperature values. No structural phase transition is observed on decreasing the temperature down to 11 K. However, the temperature dependence of the lattice parameters \emph{a(T)} and \emph{c(T)} show dramatic differences below the magnetic ordering temperature. Below $T_{C}$, \emph{a(T)} decreases continuously down to 11 K, whereas \emph{c(T)} shows a region of negative thermal expansion. This behavior shows the presence of magneto-elastic coupling. This coupling is also evident in the temperature dependence of elastic constants as discussed below.

Resonant ultrasound spectroscopy (RUS) measurements were conducted to obtain the temperature dependence of the longitudinal (\emph{C$_{11}$}) and shear (\emph{C$_{44}$}) elastic moduli. \emph{C$_{11}$} and \emph{C$_{44}$} were used to obtain the shear and longitudinal sound velocities. For polycrystalline samples, the shear velocity is given by\cite{Edward1973} v$_{s}$ = $\sqrt{\frac{C_{44}}{d}}$ and the longitudinal velocity is given by v$_{l}$ = $\sqrt{\frac{C_{11}}{d}}$, where \emph{$d$} is the density of the sample. Anderson's formula \cite{Anderson1963} was then used to calculate the Debye temperature from the average sound velocity just above the transition temperature (90 K). The Debye temperature obtained is 340 K, which is consistent with the Debye temperature estimated from $C_{p}$(T) above.

\begin{figure}[h]
\begin{center}
\includegraphics[scale=2]{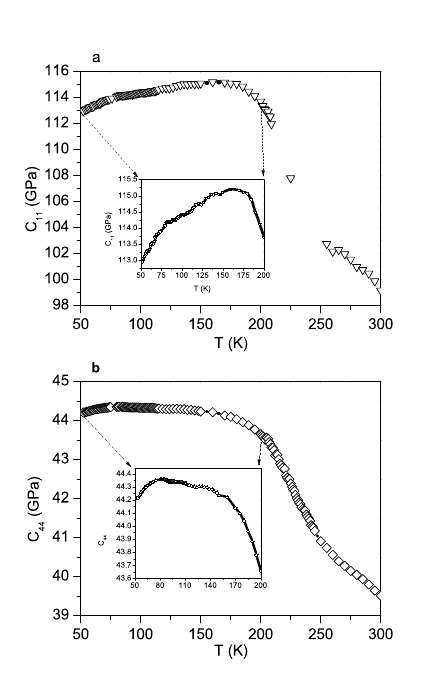}
\caption{\label{10}Variation of elastic moduli (a) $C_{11}$ and (b) $C_{44}$ as a function of temperature.}
\end{center}
\end{figure}
The temperature dependence of the longitudinal \emph{C$_{11}$} and shear \emph{C$_{44}$} elastic moduli of Cr$_{11}$Ge$_{19}$ is plotted in Figs. \ref{10}(a) and \ref{10}(b), respectively. In both figures the insets focus on the region from about 50 to 200 K so as to show the behavior near the Curie temperature. Between 220 and 250 K the ultrasonic absorption in the sample became so great that for several temperatures not enough resonances were observed to allow for an accurate determination of both elastic moduli. However, a few resonances that depend almost exclusively on \emph{C$_{44}$} remained visible throughout this region, which allowed us to follow the shear modulus over the entire temperature range. The typical temperature dependence of elastic moduli is that at higher temperatures they increase linearly with decrease in temperature and approach absolute zero with zero slope.\cite{Varshni1970,Lakkad1971} In Cr$_{11}$Ge$_{19}$, deviation from the normal behavior is observed in both the longitudinal and shear elastic constants. \emph{C$_{11}$} starts softening well above the Curie temperature without showing any remarkable feature at the transition temperature. \emph{C$_{44}$}, on the other hand, increases with decreasing temperature down to $T_{C}$ and then softens upon further cooling. This demonstrates the interaction between the magnetic ordering and the crystal lattice in Cr$_{11}$Ge$_{19}$. This behavior is reminiscent of the Invar effect in ferromagnetic materials and is in accord with the prediction of Landau's theory of second order magneto-elastic coupling. \cite{Kawald1994}

\subsection{Electronic structure calculations}

\begin{figure}[h]
\begin{center}
\includegraphics[scale=1.8]{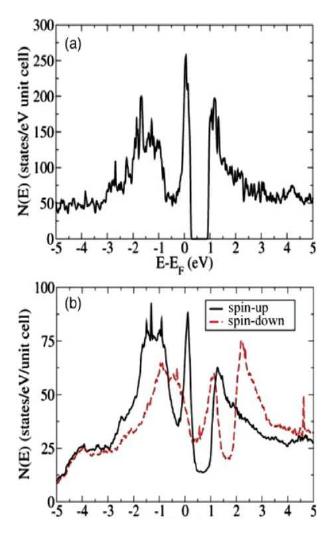}
\caption{\label{11}(Color online) Electronic density of states of Cr$_{11}$Ge$_{19}$ in (a) nonmagnetic state and (b) magnetic state.}
\end{center}
\end{figure}
Experimental results have indicated that some of the properties of Cr$_{11}$Ge$_{19}$ deviate from what is expected in an ordinary ferromagnet. In an attempt to understand these unusual behaviors we have performed first principles calculations of Cr$_{11}$Ge$_{19}$ in both a nonmagnetic state and a collinear ferromagnetic state, using the all-electron code WIEN2K \cite{wien} in the generalized gradient approximation (GGA) of Perdew, Burke, and Ernzerhof.\cite{Perdew1996} Atomic sphere radii of 2.41 and 2.13 Bohr radii were used for the Cr and Ge atoms, respectively, and an RK$_{max}$ of 7, where R is the minimum atomic sphere radius and K is the largest plane-wave vector used in the expansion.  Calculations proceeded slowly due to the large number (1680) of electrons in the unit cell; final results were converged to within 2 meV per unit cell, a small value considering the large unit cell. We find strong evidence for a magnetic ground state of Cr$_{11}$Ge$_{19}$, with the ferromagnetic ordering some 87 meV per Cr atom lower in energy than the nonmagnetic ground state.  The ordered moment averages approximately 1 $\mu_{B}$ per Cr atom, but is unevenly distributed amongst the 12 inequivalent Cr sites in the unit cell, with moment per site ranging from 0.3 to 1.7 $\mu_{B}$, which we interpret  as indirect evidence of the propensity of the system towards a noncollinear ground state. This interpretation of a noncollinear ground state is based upon the DFT constraining of the moments to be collinear, which will necessarily reduce the value of the calculated moment for those spins which in the actual physical system are not collinear.  This interpretation is strengthened by the fact that the reported \cite{Vollenkle1967} Cr-Cr nearest neighbor distances vary only from 3.124 to 3.138 \AA\ amongst the inequivalent Cr sites, so that the differences in calculated moment are more likely to be an artifact of the collinearity assumed rather than indicative of physically distinct moment values.  The calculated average moment is somewhat higher than the observed experimental value of 0.5 $\mu_{B}$; this overestimation sometimes occurs with the GGA.  Due to the time-consuming nature of the calculations we have not carried out additional local density approximation (LDA) calculations, which may better match the actual ordered moment.

To better understand the electronic structure we have calculated the electronic density of states (DOS) in both the magnetic and nonmagnetic states, using the first principles calculated band structure with approximately 300 $k$ points in the full Brillouin zone. These DOS are depicted in Fig. \ref{11}.  The nonmagnetic state [Fig. \ref{11}(a)] shows a huge peak in the density of states exactly at the Fermi energy, highly favorable towards a Stoner-type ferromagnetic instability (recall the Stoner criterion $IN_{0} > 1$, where I is the exchange correlation integral and N$_{0}$ the Fermi level density of states).  This DOS is in rough agreement with the non-self-consistent band structure calculations of Ref. \onlinecite{Pecheur1997}.   With I for Cr taken from Ref. \onlinecite{Janak1977} as 0.38 eV and the Fermi level DOS of approximately 5.7/Cr/eV, the Stoner criterion is well satisfied, and as described earlier this fits with the magnetic ground state we find.

We turn now to the magnetic state DOS [Fig. \ref{11}(b)].  The majority spin-up DOS still has a peak very near $E_{F}$, but this peak is much lower than in the nonmagnetic case. Substantial spectral weight for the spin-up states is transferred below E$_{F}$, as is expected for the majority spin, while the spin-down DOS is somewhat more equally distributed above and below E$_{F}$. We note also that the strongly magnetic nature of this system is paralleled by the spin-up and spin-down DOS not coinciding until several eV from the Fermi level. We note also that the band gap has disappeared, with instead a deep minimum in the spin-up DOS just above E$_{F}$ and a somewhat less deep minimum in the spin-down DOS.  This band gap absence, to be compared with its existence in the nonmagnetic state, is again indicative of the strong magnetism present in this material.

\section{Conclusion}

The results presented here, both from experiment and first principles calculations, indicate unusual magnetism in Cr$_{11}$Ge$_{19}$. The behavior of the magnetization and heat capacity suggest itinerant, noncollinear ferromagnetism with a Curie temperature near 88 K, and this description is supported by first principles calculations. The magnetism appears to be strongly coupled to the crystal lattice, as indicated by anomalous behavior of the lattice parameters and the elastic moduli at and below $T_C$. The influence of spin-wave excitations is observed in the heat capacity at low temperature. Interestingly, some of the properties are similar to those of MnSi and other itinerant ferromagnets. It is interesting to speculate about possible helimagnetism in Cr$_{11}$Ge$_{19}$, based on the observed properties and the nature of the crystal structure. However, the present data cannot confirm the magnetic structure, and single crystals of suitable size for neutron diffraction are not yet available. Our observations clearly point to complex and interesting magnetism in this compound, and show that further study would be of interest.

\section*{ACKNOWLEDGEMENTS}

We thank S. E. Nagler for stimulating discussions and U. K. R\"{o}$\ss$ler for pointing out that magnetic vortices have been predicted to form in the space group to which Cr$_{11}$Ge$_{19}$ belongs. Research was supported by the U.S. Department of Energy, Office of Basic Energy Sciences, Materials Sciences and Engineering Division. M.A.M. acknowledges support from the U.S. Department of Energy, Energy Efficiency and Renewable Energy, Office of Vehicle Technologies, Propulsion Materials Program. D.P. supported by the ORNL LDRD SEED funding project S12-006, ``Rare Earth Free Magnets: Compute, Create, Characterize". V.K. and M.K. acknowledge support from DOD DEPSCoR Grant No. N00014-08-1-0783 and NSF-DMR-0804719. We are grateful for the technical assistance of Douglas E. Fielden at the University of Tennessee.

\bibliographystyle{apsrev}

%\bibliography{Cr11Ge19_paper_writing}

\begin{thebibliography}{44}
\expandafter\ifx\csname natexlab\endcsname\relax\def\natexlab#1{#1}\fi
\expandafter\ifx\csname bibnamefont\endcsname\relax
  \def\bibnamefont#1{#1}\fi
\expandafter\ifx\csname bibfnamefont\endcsname\relax
  \def\bibfnamefont#1{#1}\fi
\expandafter\ifx\csname citenamefont\endcsname\relax
  \def\citenamefont#1{#1}\fi
\expandafter\ifx\csname url\endcsname\relax
  \def\url#1{\texttt{#1}}\fi
\expandafter\ifx\csname urlprefix\endcsname\relax\def\urlprefix{URL }\fi
\providecommand{\bibinfo}[2]{#2}
\providecommand{\eprint}[2][]{\url{#2}}

\bibitem[{\citenamefont{Dzaloshinsky}(1958)}]{Dzyaloshinsky1958}
\bibinfo{author}{\bibfnamefont{I.}~\bibnamefont{Dzaloshinsky}},
  \bibinfo{journal}{J. Phys. Chem. Solids} \textbf{\bibinfo{volume}{4}},
  \bibinfo{pages}{241} (\bibinfo{year}{1958}).

\bibitem[{\citenamefont{Moriya}(1960)}]{Moriya1960}
\bibinfo{author}{\bibfnamefont{T.}~\bibnamefont{Moriya}},
  \bibinfo{journal}{Phys. Rev.} \textbf{\bibinfo{volume}{120}},
  \bibinfo{pages}{91} (\bibinfo{year}{1960}).

\bibitem[{\citenamefont{Bak and Jensen}(1980)}]{Bak1980a}
\bibinfo{author}{\bibfnamefont{P.}~\bibnamefont{Bak}} \bibnamefont{and}
  \bibinfo{author}{\bibfnamefont{M.~H.} \bibnamefont{Jensen}},
  \bibinfo{journal}{J. Phys. C: Solid St. Phys.} \textbf{\bibinfo{volume}{13}},
  \bibinfo{pages}{L881} (\bibinfo{year}{1980}).

\bibitem[{\citenamefont{Nakanishi et~al.}(1980)\citenamefont{Nakanishi, Yanase,
  Hasegawa, and Kataoka}}]{Nakanishi1980b}
\bibinfo{author}{\bibfnamefont{O.}~\bibnamefont{Nakanishi}},
  \bibinfo{author}{\bibfnamefont{A.}~\bibnamefont{Yanase}},
  \bibinfo{author}{\bibfnamefont{A.}~\bibnamefont{Hasegawa}}, \bibnamefont{and}
  \bibinfo{author}{\bibfnamefont{M.}~\bibnamefont{Kataoka}},
  \bibinfo{journal}{Solid State Commun.} \textbf{\bibinfo{volume}{35}},
  \bibinfo{pages}{995} (\bibinfo{year}{1980}).

\bibitem[{\citenamefont{Fischer et~al.}(2008)\citenamefont{Fischer, Shah, and
  Rosch}}]{Fischer2008a}
\bibinfo{author}{\bibfnamefont{I.}~\bibnamefont{Fischer}},
  \bibinfo{author}{\bibfnamefont{N.}~\bibnamefont{Shah}}, \bibnamefont{and}
  \bibinfo{author}{\bibfnamefont{A.}~\bibnamefont{Rosch}},
  \bibinfo{journal}{Phys. Rev. B} \textbf{\bibinfo{volume}{77}},
  \bibinfo{pages}{024415} (\bibinfo{year}{2008}).

\bibitem[{\citenamefont{M\"{u}hlbauer et~al.}(2009)\citenamefont{M\"{u}hlbauer,
  Binz, Jonietz, Pfleiderer, Rosch, Neubauer, Georgii, and
  B\"{o}ni}}]{Muhlbauer2009}
\bibinfo{author}{\bibfnamefont{S.}~\bibnamefont{M\"{u}hlbauer}},
  \bibinfo{author}{\bibfnamefont{B.}~\bibnamefont{Binz}},
  \bibinfo{author}{\bibfnamefont{F.}~\bibnamefont{Jonietz}},
  \bibinfo{author}{\bibfnamefont{C.}~\bibnamefont{Pfleiderer}},
  \bibinfo{author}{\bibfnamefont{A.}~\bibnamefont{Rosch}},
  \bibinfo{author}{\bibfnamefont{A.}~\bibnamefont{Neubauer}},
  \bibinfo{author}{\bibfnamefont{R.}~\bibnamefont{Georgii}}, \bibnamefont{and}
  \bibinfo{author}{\bibfnamefont{P.}~\bibnamefont{B\"{o}ni}},
  \bibinfo{journal}{Science} \textbf{\bibinfo{volume}{323}},
  \bibinfo{pages}{915} (\bibinfo{year}{2009}).

\bibitem[{\citenamefont{Ho et~al.}(2010)\citenamefont{Ho, Kirkpatrick, Sang,
  and Belitz}}]{Ho2010b}
\bibinfo{author}{\bibfnamefont{K.Y.} \bibnamefont{Ho}},
  \bibinfo{author}{\bibfnamefont{T.R.}~\bibnamefont{Kirkpatrick}},
  \bibinfo{author}{\bibfnamefont{Y.}~\bibnamefont{Sang}}, \bibnamefont{and}
  \bibinfo{author}{\bibfnamefont{D.}~\bibnamefont{Belitz}},
  \bibinfo{journal}{Phys. Rev. B} \textbf{\bibinfo{volume}{82}},
  \bibinfo{pages}{134427} (\bibinfo{year}{2010}).

\bibitem[{\citenamefont{Pfleiderer et~al.}(2009)\citenamefont{Pfleiderer,
  Neubauer, M\"{u}hlbauer, Jonietz, Janoschek, Legl, Ritz, M\"{u}nzer, Franz,
  Niklowitz et~al.}}]{Pfleiderer2009a}
\bibinfo{author}{\bibfnamefont{C.}~\bibnamefont{Pfleiderer}},
  \bibinfo{author}{\bibfnamefont{A.}~\bibnamefont{Neubauer}},
  \bibinfo{author}{\bibfnamefont{S.}~\bibnamefont{M\"{u}hlbauer}},
  \bibinfo{author}{\bibfnamefont{F.}~\bibnamefont{Jonietz}},
  \bibinfo{author}{\bibfnamefont{M.}~\bibnamefont{Janoschek}},
  \bibinfo{author}{\bibfnamefont{S.}~\bibnamefont{Legl}},
  \bibinfo{author}{\bibfnamefont{R.}~\bibnamefont{Ritz}},
  \bibinfo{author}{\bibfnamefont{W.}~\bibnamefont{M\"{u}nzer}},
  \bibinfo{author}{\bibfnamefont{C.}~\bibnamefont{Franz}},
  \bibinfo{author}{\bibfnamefont{P.~G.} \bibnamefont{Niklowitz}},
  \bibnamefont{et~al.}, \bibinfo{journal}{J. Phys.: Condens. Matter}
  \textbf{\bibinfo{volume}{21}}, \bibinfo{pages}{164215}
  (\bibinfo{year}{2009}).

\bibitem[{\citenamefont{Uchida et~al.}(2006)\citenamefont{Uchida, Onose,
  Matsui, and Tokura}}]{Uchida2006b}
\bibinfo{author}{\bibfnamefont{M.}~\bibnamefont{Uchida}},
  \bibinfo{author}{\bibfnamefont{Y.}~\bibnamefont{Onose}},
  \bibinfo{author}{\bibfnamefont{Y.}~\bibnamefont{Matsui}}, \bibnamefont{and}
  \bibinfo{author}{\bibfnamefont{Y.}~\bibnamefont{Tokura}},
  \bibinfo{journal}{Science} \textbf{\bibinfo{volume}{311}},
  \bibinfo{pages}{359} (\bibinfo{year}{2006}).

\bibitem[{\citenamefont{M\"{u}nzer et~al.}(2010)\citenamefont{M\"{u}nzer,
  Neubauer, Adams, M\"{u}hlbauer, Franz, Jonietz, Georgii, B\"{o}ni, Pedersen,
  Schmidt et~al.}}]{Munzer2010}
\bibinfo{author}{\bibfnamefont{W.}~\bibnamefont{M\"{u}nzer}},
  \bibinfo{author}{\bibfnamefont{A.}~\bibnamefont{Neubauer}},
  \bibinfo{author}{\bibfnamefont{T.}~\bibnamefont{Adams}},
  \bibinfo{author}{\bibfnamefont{S.}~\bibnamefont{M\"{u}hlbauer}},
  \bibinfo{author}{\bibfnamefont{C.}~\bibnamefont{Franz}},
  \bibinfo{author}{\bibfnamefont{F.}~\bibnamefont{Jonietz}},
  \bibinfo{author}{\bibfnamefont{R.}~\bibnamefont{Georgii}},
  \bibinfo{author}{\bibfnamefont{P.}~\bibnamefont{B\"{o}ni}},
  \bibinfo{author}{\bibfnamefont{B.}~\bibnamefont{Pedersen}},
  \bibinfo{author}{\bibfnamefont{M.}~\bibnamefont{Schmidt}},
  \bibnamefont{et~al.}, \bibinfo{journal}{Phys. Rev. B}
  \textbf{\bibinfo{volume}{81}}, \bibinfo{pages}{041203(R)}
  (\bibinfo{year}{2010}).

\bibitem[{\citenamefont{Miyadai et~al.}(1983)\citenamefont{Miyadai, Kikuchi,
  Kondo, Sakka, Arai, and Ishikawa}}]{Miyadai1983}
\bibinfo{author}{\bibfnamefont{T.}~\bibnamefont{Miyadai}},
  \bibinfo{author}{\bibfnamefont{K.}~\bibnamefont{Kikuchi}},
  \bibinfo{author}{\bibfnamefont{H.}~\bibnamefont{Kondo}},
  \bibinfo{author}{\bibfnamefont{S.}~\bibnamefont{Sakka}},
  \bibinfo{author}{\bibfnamefont{K.}~\bibnamefont{Arai}}, \bibnamefont{and}
  \bibinfo{author}{\bibfnamefont{Y.}~\bibnamefont{Ishikawa}},
  \bibinfo{journal}{J. Phys. Soc. Japan} \textbf{\bibinfo{volume}{52}},
  \bibinfo{pages}{1394} (\bibinfo{year}{1983}).

\bibitem[{\citenamefont{Moriya and Miyadai}(1982)}]{Moriya1982}
\bibinfo{author}{\bibfnamefont{T.}~\bibnamefont{Moriya}} \bibnamefont{and}
  \bibinfo{author}{\bibfnamefont{T.}~\bibnamefont{Miyadai}},
  \bibinfo{journal}{Solid State Commun.} \textbf{\bibinfo{volume}{42}},
  \bibinfo{pages}{209} (\bibinfo{year}{1982}).

\bibitem[{\citenamefont{Togawa et~al.}(2012)\citenamefont{Togawa, Koyama,
  Takayanagi, Mori, Kousaka, Akimitsu, Nishihara, Inoue, Ovchinnikov, and
  Kishine}}]{Togawa2012}
\bibinfo{author}{\bibfnamefont{Y.}~\bibnamefont{Togawa}},
  \bibinfo{author}{\bibfnamefont{T.}~\bibnamefont{Koyama}},
  \bibinfo{author}{\bibfnamefont{K.}~\bibnamefont{Takayanagi}},
  \bibinfo{author}{\bibfnamefont{S.}~\bibnamefont{Mori}},
  \bibinfo{author}{\bibfnamefont{Y.}~\bibnamefont{Kousaka}},
  \bibinfo{author}{\bibfnamefont{J.}~\bibnamefont{Akimitsu}},
  \bibinfo{author}{\bibfnamefont{S.}~\bibnamefont{Nishihara}},
  \bibinfo{author}{\bibfnamefont{K.}~\bibnamefont{Inoue}},
  \bibinfo{author}{\bibfnamefont{A.~S.}~\bibnamefont{Ovchinnikov}},
  \bibnamefont{and} \bibinfo{author}{\bibfnamefont{J.}~\bibnamefont{Kishine}},
  \bibinfo{journal}{Phys. Rev. Lett.} \textbf{\bibinfo{volume}{108}},
  \bibinfo{pages}{107202} (\bibinfo{year}{2012}).
 \bibitem[{\citenamefont{Vollenkle et~al.}(1967)}]{Vollenkle1967}
 \bibinfo{author}{\bibfnamefont{H.}~\bibnamefont{V\"{o}llenkle}},
  \bibinfo{author}{\bibfnamefont{A.}~\bibnamefont{Preisinger}},
  \bibinfo{author}{\bibfnamefont{H.}~\bibnamefont{Nowotny}}, \bibnamefont{and}
  \bibinfo{author}{\bibfnamefont{A.}~\bibnamefont{Wittmann}},
  \bibinfo{journal}{Z. Kristallogr.} \textbf{\bibinfo{volume}{124}},
  \bibinfo{pages}{9} (\bibinfo{year}{1967}).

\bibitem[{\citenamefont{Zagryazhskii et~al.}(1968)\citenamefont{Zagryazhskii,
  Gel'd, and Shtol'ts}}]{Zagryazhskii1968}
\bibinfo{author}{\bibfnamefont{V.~L.} \bibnamefont{Zagryazhskii}},
  \bibinfo{author}{\bibfnamefont{P.~V.} \bibnamefont{Gel'd}}, \bibnamefont{and}
  \bibinfo{author}{\bibfnamefont{A.~K.} \bibnamefont{Shtol'ts}},
  \bibinfo{journal}{Soviet Physics Journal} \textbf{\bibinfo{volume}{11}},
  \bibinfo{pages}{23} (\bibinfo{year}{1968}).

\bibitem[{\citenamefont{Kolenda et~al.}(1980)\citenamefont{Kolenda, Stoch, and
  Szytula}}]{Kolenda1980}
\bibinfo{author}{\bibfnamefont{M.}~\bibnamefont{Kolenda}},
  \bibinfo{author}{\bibfnamefont{J.}~\bibnamefont{Stoch}}, \bibnamefont{and}
  \bibinfo{author}{\bibfnamefont{A.}~\bibnamefont{Szytula}},
  \bibinfo{journal}{J. Magn. Magn. Mater} \textbf{\bibinfo{volume}{20}},
  \bibinfo{pages}{99} (\bibinfo{year}{1980}).

  \bibitem[{\citenamefont{Bogdanov and Yablonskii}(1989)\citenamefont{Bogdanov and
  Yablonskii}}]{Bogdanov1989a}
\bibinfo{author}{\bibfnamefont{A.~N.}~\bibnamefont{Bogdanov}}, \bibnamefont{and}
\bibinfo{author}{\bibfnamefont{D.~A.}~\bibnamefont{Yablonskii}},
\bibinfo{journal}{Sov. Phys. JEPT} \textbf{\bibinfo{volume}{68}},
\bibinfo{pages}{101} (\bibinfo{year}{1989}).

\bibitem[{\citenamefont{Bogdanov et~al.}(1989)\citenamefont{Bogdanov, Kudinov and
  Yablonskii}}]{Bogdanov1989b}
\bibinfo{author}{\bibfnamefont{A.~N.}~\bibnamefont{Bogdanov}},
  \bibinfo{author}{\bibfnamefont{M.~V.}~\bibnamefont{Kudinov}}, \bibnamefont{and}
  \bibinfo{author}{\bibfnamefont{D.~A.}~\bibnamefont{Yablonskii}},
  \bibinfo{journal}{Sov. Phys. Solid State} \textbf{\bibinfo{volume}{31}},
  \bibinfo{pages}{1707} (\bibinfo{year}{1989}).

\bibitem[{\citenamefont{Bogdanov and Hubert}(1994)\citenamefont{Kolenda, Stoch, and
  Szytula}}]{Bogdanov1994}
\bibinfo{author}{\bibfnamefont{A.}~\bibnamefont{Bogdanov}}, \bibnamefont{and}
  \bibinfo{author}{\bibfnamefont{A.}~\bibnamefont{Hubert}},
  \bibinfo{journal}{J. Magn. Magn. Mater} \textbf{\bibinfo{volume}{138}},
  \bibinfo{pages}{255} (\bibinfo{year}{1994}).


\bibitem[{\citenamefont{P\'{e}cheur et~al.}(1997)\citenamefont{P\'{e}cheur,
  Toussaint, Kenzari, Malaman, and Welter}}]{Pecheur1997}
\bibinfo{author}{\bibfnamefont{P.}~\bibnamefont{P\'{e}cheur}},
  \bibinfo{author}{\bibfnamefont{G.}~\bibnamefont{Toussaint}},
  \bibinfo{author}{\bibfnamefont{H.}~\bibnamefont{Kenzari}},
  \bibinfo{author}{\bibfnamefont{B.}~\bibnamefont{Malaman}}, \bibnamefont{and}
  \bibinfo{author}{\bibfnamefont{R.}~\bibnamefont{Welter}},
  \bibinfo{journal}{J. Alloy. Compd.} \textbf{\bibinfo{volume}{262-263}},
  \bibinfo{pages}{363} (\bibinfo{year}{1997}).

\bibitem[{\citenamefont{Caillat et~al.}(1997)\citenamefont{Caillat, Fleurial,
  and Borshchevsky}}]{Caillat1997}
\bibinfo{author}{\bibfnamefont{T.}~\bibnamefont{Caillat}},
  \bibinfo{author}{\bibfnamefont{J.-P.} \bibnamefont{Fleurial}},
  \bibnamefont{and}
  \bibinfo{author}{\bibfnamefont{A.}~\bibnamefont{Borshchevsky}},
  \bibinfo{journal}{J. Alloy. Compd.} \textbf{\bibinfo{volume}{252}},
  \bibinfo{pages}{12} (\bibinfo{year}{1997}).


\bibitem[{\citenamefont{Migliori and Maynard}(2005)}]{Migliori2005}
\bibinfo{author}{\bibfnamefont{A.}~\bibnamefont{Migliori}} \bibnamefont{and}
  \bibinfo{author}{\bibfnamefont{J.~D.} \bibnamefont{Maynard}},
  \bibinfo{journal}{Rev. Sci. Instrum.} \textbf{\bibinfo{volume}{76}},
  \bibinfo{pages}{121301} (\bibinfo{year}{2005}).

\bibitem[{\citenamefont{Fredrickson et~al.}(2004)\citenamefont{Fredrickson,
  Lee, and Hoffmann}}]{Fredrickson2004a}
\bibinfo{author}{\bibfnamefont{D.~C.} \bibnamefont{Fredrickson}},
  \bibinfo{author}{\bibfnamefont{S.}~\bibnamefont{Lee}}, \bibnamefont{and}
  \bibinfo{author}{\bibfnamefont{R.}~\bibnamefont{Hoffmann}},
  \bibinfo{journal}{Inorg. Chem.} \textbf{\bibinfo{volume}{43}},
  \bibinfo{pages}{6159} (\bibinfo{year}{2004}).

\bibitem[{\citenamefont{Wohlfarth}(1977)}]{Wohlfarth1977}
\bibinfo{author}{\bibfnamefont{E.}~\bibnamefont{Wohlfarth}},
  \bibinfo{journal}{Physica B} \textbf{\bibinfo{volume}{91}},
  \bibinfo{pages}{305} (\bibinfo{year}{1977}).

\bibitem[{\citenamefont{Wohlfarth}(1968)}]{Wohlfarth1968}
\bibinfo{author}{\bibfnamefont{E.~P.} \bibnamefont{Wohlfarth}},
  \bibinfo{journal}{J. Appl. Phys.} \textbf{\bibinfo{volume}{39}},
  \bibinfo{pages}{1061} (\bibinfo{year}{1968}).

\bibitem[{\citenamefont{Edwards and Wohlfarth}(1968)}]{Edwards1968}
\bibinfo{author}{\bibfnamefont{D.~M.} \bibnamefont{Edwards}} \bibnamefont{and}
  \bibinfo{author}{\bibfnamefont{E.~P.} \bibnamefont{Wohlfarth}},
  \bibinfo{journal}{Proc. Roy. Soc. A.} \textbf{\bibinfo{volume}{303}},
  \bibinfo{pages}{127} (\bibinfo{year}{1968}).

\bibitem[{\citenamefont{Arrott and Noakes}(1967)}]{Arrott1967}
\bibinfo{author}{\bibfnamefont{A.}~\bibnamefont{Arrott}} \bibnamefont{and}
  \bibinfo{author}{\bibfnamefont{J.~E.}~\bibnamefont{Noakes}},
  \bibinfo{journal}{Phys. Rev. Lett.} \textbf{\bibinfo{volume}{19}},
  \bibinfo{pages}{786} (\bibinfo{year}{1967}).

\bibitem[{\citenamefont{Chattopadhyay et~al.}(2009)\citenamefont{Chattopadhyay,
  Arora, and Roy}}]{Chattopadhyay2009}
\bibinfo{author}{\bibfnamefont{M.~K.} \bibnamefont{Chattopadhyay}},
  \bibinfo{author}{\bibfnamefont{P.}~\bibnamefont{Arora}}, \bibnamefont{and}
  \bibinfo{author}{\bibfnamefont{S.~B.} \bibnamefont{Roy}},
  \bibinfo{journal}{J. Phys.: Condens. Matter.} \textbf{\bibinfo{volume}{21}},
  \bibinfo{pages}{296003} (\bibinfo{year}{2009}).

\bibitem[{\citenamefont{Ohta and Yoshimura}(2009)}]{Ohta2009}
\bibinfo{author}{\bibfnamefont{H.}~\bibnamefont{Ohta}} \bibnamefont{and}
  \bibinfo{author}{\bibfnamefont{K.}~\bibnamefont{Yoshimura}},
  \bibinfo{journal}{Phys. Rev. B} \textbf{\bibinfo{volume}{79}},
  \bibinfo{pages}{184407} (\bibinfo{year}{2009}).

\bibitem[{\citenamefont{Takahashi}(1986)}]{Takahashi1986}
\bibinfo{author}{\bibfnamefont{Y.}~\bibnamefont{Takahashi}},
  \bibinfo{journal}{J. Phys. Soc. Japan} \textbf{\bibinfo{volume}{55}},
  \bibinfo{pages}{3553} (\bibinfo{year}{1986}).

\bibitem[{\citenamefont{Shimizu et~al.}(1990)\citenamefont{Shimizu, Maruyama,
  Yamazaki, and Watanabe}}]{Shimizu1990}
\bibinfo{author}{\bibfnamefont{K.}~\bibnamefont{Shimizu}},
  \bibinfo{author}{\bibfnamefont{H.}~\bibnamefont{Maruyama}},
  \bibinfo{author}{\bibfnamefont{H.}~\bibnamefont{Yamazaki}}, \bibnamefont{and}
  \bibinfo{author}{\bibfnamefont{H.}~\bibnamefont{Watanabe}},
  \bibinfo{journal}{J. Phys. Soc. Japan} \textbf{\bibinfo{volume}{59}},
  \bibinfo{pages}{305} (\bibinfo{year}{1990}).

\bibitem[{\citenamefont{Ho et~al.}(1981)\citenamefont{Ho, Maartense, and
  Williams}}]{SCHo1981}
\bibinfo{author}{\bibfnamefont{S.~C.} \bibnamefont{Ho}},
  \bibinfo{author}{\bibfnamefont{I.}~\bibnamefont{Maartense}},
  \bibnamefont{and} \bibinfo{author}{\bibfnamefont{G.}~\bibnamefont{Williams}},
  \bibinfo{journal}{J. Phys. F: Metal Phys.} \textbf{\bibinfo{volume}{11}},
  \bibinfo{pages}{699} (\bibinfo{year}{1981}).

\bibitem[{\citenamefont{Vannette et~al.}(2008)\citenamefont{Vannette, Sefat,
  Jia, Law, Lapertot, Bud’ko, Canfield, Schmalian, and
  Prozorov}}]{Vannette2008b}
\bibinfo{author}{\bibfnamefont{M.}~\bibnamefont{Vannette}},
  \bibinfo{author}{\bibfnamefont{A.~S.}~\bibnamefont{Sefat}},
  \bibinfo{author}{\bibfnamefont{S.}~\bibnamefont{Jia}},
  \bibinfo{author}{\bibfnamefont{S.~A.}~\bibnamefont{Law}},
  \bibinfo{author}{\bibfnamefont{G.}~\bibnamefont{Lapertot}},
  \bibinfo{author}{\bibfnamefont{S.~L.}~\bibnamefont{Bud'ko}},
  \bibinfo{author}{\bibfnamefont{P.~C.}~\bibnamefont{Canfield}},
  \bibinfo{author}{\bibfnamefont{J.}~\bibnamefont{Schmalian}},
  \bibnamefont{and} \bibinfo{author}{\bibfnamefont{R.}~\bibnamefont{Prozorov}},
  \bibinfo{journal}{J. Magn. Magn. Mater} \textbf{\bibinfo{volume}{320}},
  \bibinfo{pages}{354} (\bibinfo{year}{2008}).

\bibitem[{\citenamefont{Thessieu et~al.}(1997)\citenamefont{Thessieu,
  Pfleiderer, Stepanov, and Flouquet}}]{Thessieu1997a}
\bibinfo{author}{\bibfnamefont{C.}~\bibnamefont{Thessieu}},
  \bibinfo{author}{\bibfnamefont{C.}~\bibnamefont{Pfleiderer}},
  \bibinfo{author}{\bibfnamefont{A.~N.} \bibnamefont{Stepanov}},
  \bibnamefont{and} \bibinfo{author}{\bibfnamefont{J.}~\bibnamefont{Flouquet}},
  \bibinfo{journal}{J. Phys.: Condens. Matter}
  \textbf{\bibinfo{volume}{9}}, \bibinfo{pages}{6677}
  (\bibinfo{year}{1997}).

\bibitem[{\citenamefont{Wilhelm et~al.}(2011)\citenamefont{Wilhelm, Baenitz,
  Schmidt, R\"{o}{\ss}~ler, Leonov, and Bogdanov}}]{Wilhelm2011}
\bibinfo{author}{\bibfnamefont{H.}~\bibnamefont{Wilhelm}},
  \bibinfo{author}{\bibfnamefont{M.}~\bibnamefont{Baenitz}},
  \bibinfo{author}{\bibfnamefont{M.}~\bibnamefont{Schmidt}},
  \bibinfo{author}{\bibfnamefont{U.~K.}~\bibnamefont{R\"{o}$\ss$ler}},
  \bibinfo{author}{\bibfnamefont{A.~A.}~\bibnamefont{Leonov}}, \bibnamefont{and}
  \bibinfo{author}{\bibfnamefont{A.~N.}~\bibnamefont{Bogdanov}},
  \bibinfo{journal}{Phys. Rev. Lett.} \textbf{\bibinfo{volume}{107}},
  \bibinfo{pages}{127203} (\bibinfo{year}{2011}).

\bibitem[{\citenamefont{Mohn and Hilscher}(1989)}]{Mohn1989}
\bibinfo{author}{\bibfnamefont{P.}~\bibnamefont{Mohn}} \bibnamefont{and}
  \bibinfo{author}{\bibfnamefont{G.}~\bibnamefont{Hilscher}},
  \bibinfo{journal}{Phys. Rev. B} \textbf{\bibinfo{volume}{40}},
  \bibinfo{pages}{9126} (\bibinfo{year}{1989}).

\bibitem[{\citenamefont{Huber et~al.}(1975)\citenamefont{Huber, Maple, and
  Wohlleben}}]{Huber1975}
\bibinfo{author}{\bibfnamefont{J.~G.} \bibnamefont{Huber}},
  \bibinfo{author}{\bibfnamefont{M.~B.} \bibnamefont{Maple}}, \bibnamefont{and}
  \bibinfo{author}{\bibfnamefont{D.}~\bibnamefont{Wohlleben}},
  \bibinfo{journal}{Solid State Commun.} \textbf{\bibinfo{volume}{16}},
  \bibinfo{pages}{211} (\bibinfo{year}{1975}).

\bibitem[{\citenamefont{Ogawa and Sakamoto}(1967)}]{Ogawa1967}
\bibinfo{author}{\bibfnamefont{S.}~\bibnamefont{Ogawa}} \bibnamefont{and}
  \bibinfo{author}{\bibfnamefont{N.}~\bibnamefont{Sakamoto}},
  \bibinfo{journal}{J. Phys. Soc. Japan} \textbf{\bibinfo{volume}{22}},
  \bibinfo{pages}{1214} (\bibinfo{year}{1967}).

\bibitem[{\citenamefont{Mohn}(2006)\citenamefont{Mohn}}]{Mohn2006}
\bibinfo{author}{\bibfnamefont{P.}~\bibnamefont{Mohn}},
  \emph{\bibinfo{title}{{Magnetism in the Solid State}}}
  (\bibinfo{publisher}{Springer}, \bibinfo{address}{Heidelberg},
  \bibinfo{year}{2006}).

\bibitem[{\citenamefont{Schreiber et~al.}(1973)\citenamefont{Schreiber,
  Anderson, and Soga}}]{Edward1973}
\bibinfo{author}{\bibfnamefont{E.}~\bibnamefont{Schreiber}},
  \bibinfo{author}{\bibfnamefont{O.~L.} \bibnamefont{Anderson}},
  \bibnamefont{and} \bibinfo{author}{\bibfnamefont{N.}~\bibnamefont{Soga}},
  \emph{\bibinfo{title}{{Elastic constants and their measurement}}}
  (\bibinfo{publisher}{McGraw-Hill}, \bibinfo{address}{New York},
  \bibinfo{year}{1973}).

\bibitem[{\citenamefont{Anderson}(1963)}]{Anderson1963}
\bibinfo{author}{\bibfnamefont{O.~L.} \bibnamefont{Anderson}},
  \bibinfo{journal}{J. Phys. Chem. Solids} \textbf{\bibinfo{volume}{24}},
  \bibinfo{pages}{909} (\bibinfo{year}{1963}).

\bibitem[{\citenamefont{Varshni}(1970)}]{Varshni1970}
\bibinfo{author}{\bibfnamefont{Y.~P.} \bibnamefont{Varshni}},
  \bibinfo{journal}{Phys. Rev. B} \textbf{\bibinfo{volume}{2}}, \bibinfo{pages}{3952}
  (\bibinfo{year}{1970}).

\bibitem[{\citenamefont{Lakkad}(1971)}]{Lakkad1971}
\bibinfo{author}{\bibfnamefont{S.~C.} \bibnamefont{Lakkad}},
  \bibinfo{journal}{J. Appl. Phys.} \textbf{\bibinfo{volume}{42}},
  \bibinfo{pages}{4277} (\bibinfo{year}{1971}).

\bibitem[{\citenamefont{Kawald et~al.}(1994)\citenamefont{Kawald, Mitze, Bach,
  Pelzl, and Saunders}}]{Kawald1994}
\bibinfo{author}{\bibfnamefont{U.}~\bibnamefont{Kawald}},
  \bibinfo{author}{\bibfnamefont{O.}~\bibnamefont{Mitze}},
  \bibinfo{author}{\bibfnamefont{H.}~\bibnamefont{Bach}},
  \bibinfo{author}{\bibfnamefont{J.}~\bibnamefont{Pelzl}}, \bibnamefont{and}
  \bibinfo{author}{\bibfnamefont{G.~A.} \bibnamefont{Saunders}},
  \bibinfo{journal}{J. Phys.: Condens. Matter}
  \textbf{\bibinfo{volume}{6}}, \bibinfo{pages}{9697}
  (\bibinfo{year}{1994}).

\bibitem[{\citenamefont{Blaha et~al.}(2001)\citenamefont{Blaha, Schwarz,
  Madsen, Kvasnicka, and Luitz}}]{wien}
\bibinfo{author}{\bibfnamefont{P.}~\bibnamefont{Blaha}},
  \bibinfo{author}{\bibfnamefont{K.}~\bibnamefont{Schwarz}},
  \bibinfo{author}{\bibfnamefont{G.~K.~H.} \bibnamefont{Madsen}},
  \bibinfo{author}{\bibfnamefont{D.}~\bibnamefont{Kvasnicka}},
  \bibnamefont{and} \bibinfo{author}{\bibfnamefont{J.}~\bibnamefont{Luitz}},
  \emph{\bibinfo{title}{{WIEN2k, An Augmented Plane Wave + Local Orbitals
  Program for Calculating Crystal Properties}}} (\bibinfo{publisher}{Karlheinz
  Schwarz, Techn. Universitat Wein}, \bibinfo{address}{Austria},
  \bibinfo{year}{2001}).

\bibitem[{\citenamefont{Perdew et~al.}(1996)\citenamefont{Perdew, Burke, and
  Ernzerhof}}]{Perdew1996}
\bibinfo{author}{\bibfnamefont{J.~P.}~\bibnamefont{Perdew}},
  \bibinfo{author}{\bibfnamefont{K.}~\bibnamefont{Burke}}, \bibnamefont{and}
  \bibinfo{author}{\bibfnamefont{M.}~\bibnamefont{Ernzerhof}},
  \bibinfo{journal}{Phys. Rev. Lett.} \textbf{\bibinfo{volume}{77}},
  \bibinfo{pages}{3865} (\bibinfo{year}{1996}).

\bibitem[{\citenamefont{Janak}(1977)}]{Janak1977}
\bibinfo{author}{\bibfnamefont{J.~F.} \bibnamefont{Janak}},
  \bibinfo{journal}{Phys. Rev. B} \textbf{\bibinfo{volume}{16}},
  \bibinfo{pages}{255} (\bibinfo{year}{1977}).

\end{thebibliography}

\end{document}